\documentstyle[12pt,aps]{revtex}%
\newcommand{\bce}{\begin{center}}
\newcommand{\ece}{\end{center}}
\newcommand{\be}{\begin{equation}}
\newcommand{\ee}{\end{equation}}
\newcommand{\bea}{\vspace{0.25cm}\begin{eqnarray}}
\newcommand{\eea}{\end{eqnarray}}

\def\PLA{{Phys. Lett.}  A }
\def\PRL{{Phys. Rev. Lett.} }
\def\PRA{{Phys. Rev.} A }

\def\PRD{{Phys. Rev.} D }

\begin{document} \draft

\vskip 3cm

\begin{center}
{\bf {\LARGE Quantum non-demolition (QND) modulation of quantum interference}}
\end{center}

\vspace{ .25cm}
\begin{center}
{M.Genovese \footnote{ \small  genovese@ien.it}, C.Novero} 
\\[0pt]

Istituto Elettrotecnico Nazionale Galileo Ferraris \\[0pt]
Str. delle Cacce 91 \\[0pt]
I-10135 Torino, Italy
\end{center}

\vspace{ 4.cm} {\large  Abstract }
\vskip 0.7cm

We propose an experiment where quantum interference between two different
paths is modulated by means of a QND measurement on one or both the arms of
an interferometer. The QND measurement is achieved in a Kerr cell. We
illustrate a scheme for the realisation of this experiment and discuss some
further developments.    

\vspace{ 1.cm} {\large   }
\vskip 0.7cm

In studying the foundations of quantum mechanics, in recent years,
considerable interest has been  devoted to the investigation  of quantum
interference and to its disappearence when {\it welcher 
Weg} (which path) information is obtained. An interesting example of
quantum interference and of  its non-intuitive properties is provided by
the possibility of erasing (even after the measurement of the quantum
system) the { \it welcher Weg} information re-obtaining the interference:
the so called quantum eraser
(of which many different versions have been proposed \cite{QEp1,QEp2} and
realised \cite{QEexp1,QEexp2}).
These recent studies on quantum interference have highlighted how the
interference is wiped out not by momentum kinks or due to the uncertainty
principle, but by  destroying the entanglement related to the possibility of
obtaining the { \it welcher Weg} information \cite{Wwi,SM}. 
 
The great interest in understanding these fundamental aspects of quantum
mechanics justifies the proposal and realisation of new experiments which
permit a deeper study of these phenomena.

Another extremely interesting field of  investigation in the foundation of
quantum mechanics concerns the possibility of performing  measurements 
without disturbing the measured system by introducing uncontrollable
quantum fluctuations:  so called Quantum Non-Demolition (QND) measurements
\cite{QNDth,QNDexp}. 

QND measurements can also be used to obtain  { \it welcher Weg}
information. One can show that in this case, the fact of having obtained a
{ \it welcher Weg} information leads to the destruction of quantum
interference \cite{SM}.
 
In this letter we study the effect of a QND measurement on quantum
interference and in particular the effect of erasure of the {\it welcher
Weg} information obtained by this measurement. We believe that the
implementation of this experiment will provide a very clear and interesting
example of the effect on a quantum system of a {\it Welcher Weg}
measurement and of its partial or total erasure.
 
Let us consider the set-up of fig.1, a "signal" photon enters the
Beam-Splitter I (BS I) from port 1;
assuming for simplicity a 50\% BS (the treatment of  the non 50\% case is a
trivial extension) 
we will have after the BS

\be
a_2={a_1 + i a_0 \over \sqrt{2} } \, \, \, \, \, a_3={a_0 + i a_1 \over
\sqrt{2}}
\label{eq:a2a3}
\ee

Furthermore, a probe laser crosses the Kerr cell on the third arm acquiring
a phase which in principle will be measurable with  homodyne detection
providing { \it welcher Weg} information.

Assuming \cite{SM} an interaction Hamiltonian of the form:

\be
H = \omega _s n_s + \omega _p n_p + {\chi _s \over 2} n_s^2 + {\chi _p
\over 2} n_p^2 + 2 \chi n_s n_p
\label{eq:H}
\ee

(where $n_s$ and $n_p$ are the 
photon number operators for the signal and probe fields respectively,
$\omega_s$ and $\omega_p$ their respective frequencies and $\chi_s$,
$\chi_p$ and $\chi$ the non-linear coefficients), one can see that the
photon number remains unchanged, whilst the phases change in both
the signal and the probe fields. In more detail, (using the simplifying
hypothesis of having chosen an interaction time T such that
$\chi _p n_p T = 2 \pi N$ where $N$ is an integer ) one finds

\be
\exp{ 
\left [ -i \left (
 \omega_p+ \chi_p /2 + 
2 \chi n_s \right ) T \right ]
}
\label{eq:fase1p}
\ee

for the probe laser phase and 

\be
\exp{ 
\left [ -i \left (
 \omega_s + \chi_s /2 + \chi_s n_3+
2 \chi n_p \right ) T \right ]
}
\label{eq:fasep}
\ee
for the $3^{rd}$ arm signal field (here and in the following $n_i =
a_i^{\dag} a_i$).

After recombination on Beam Splitter II (BS II) we  have:

\be
a_4 = {a_2 + i a_3 \exp{ \left [ -i \left [
(\omega_s + \chi_s /2 + \chi_s n_3 + 2 \chi n_p )  T + \Theta \right ]
\right]} \over \sqrt{2}}
\label{eq:a5}
\ee

where the phase $\Theta$ takes into account different lengths of arms 2 and
3 and could be varied interposing a variable phase shift on one of the
interferometer arms.

If we assume, as a simplifying  approximation, that the probe laser is
described by a coherent field $\vert \nu \rangle$, the initial 
state is $ = \vert 1 \rangle_1 \vert 0 \rangle_0 \vert
\nu \rangle_p $, and

\be
\langle \Psi \vert n_4 \vert \Psi \rangle = 1/2 [ 1- exp [ -2 \vert \nu \vert^2
sin ^2 (\chi T) ] cos[ (\omega_s + \chi_s ) T + \Theta + \vert \nu \vert ^2 sin
(2 \chi T) ]]
\ee

To this point, the usual treatment \cite{SM} of QND ideal { \it welcher
Weg} experiment has been considered,
with the well known result that the probe laser acquires a phase which
permits the identification of the path followed by the signal photon by  a
homodyne 
detection. As the signal-to-noise ratio in the homodyne measurement is $R =4
 \vert \nu \vert \sin (\chi T)$ \cite{SM}, one finds 
a suppression of interference, whose visibility is related to $R$ and given
by $\exp{ ( - R^2 /8)}$. It is, therefore, evident that  increasing  the
signal-to-noise ratio $R$, i.e. a better determination of the probe laser
phase,  directly relates to the disappearance of interference fringes for
the signal field. Also in this case, we find that the disappearance of
interference is clearly due 
to the acquisition of { \it welcher Weg} information and not to
disturbances introduced into the system by the measurement.

A quantitative definition of {\it distinguishability D} of the ways is
given in Ref. \cite{Englert}, where it is also shown how {\it D} and the
fringe visibility {\it V } satisfy the inequality 
\be 
D^2 + V^2 \le 1 
\label{eq:dise}
\ee
 According to the discussion given in Ref. \cite{Englert}, when the initial
"detector" state can be described by a statistical operator corresponding
to a pure state and thus, in particular, by $\vert \nu \rangle \langle \nu
\vert$, the equal sign holds in Eq. \ref{eq:dise}. 

Let us now consider the insertion of a second Kerr cell on the $2^{nd}$ arm
 (KII in fig.1) and let us assume that the interaction time between the
probe and signal fields in this cell is $T'$.

We now have:

\be
a_4 = { a_2 \exp{ \left [ -i \left [
(\omega_s + \chi_s /2 + \chi_s n_2 +
2 \chi n_p )  T' \right ] \right]}
+ i a_3 \exp{ \left [ -i \left [
(\omega_s + \chi_s /2 + \chi_s n_3 +
2 \chi n_p )  T + \Theta \right ] \right]} \over \sqrt{2}}
\label{eq:a5ii}
\ee

If the distance between the two Kerr cells is less than the coherence
length of the probe field, the two paths will become indistinguishable
again ($D \rightarrow 0$) and interference will be achieved once more ($V
\rightarrow 1$). Setting
\be 
\beta = (\omega_s + \chi_s /2 + \chi_s n_s + 2 \chi n_p ) \, ,
\label{eq:beta}
\ee
one has

\be
\langle \Psi \vert n_4 \vert \Psi \rangle = 1/4 \langle \Psi \vert n_1
[2 -(exp[-i (\Theta + \beta (T-T'))]+exp[i (\Theta + \beta (T-T'))]
] \vert \Psi \rangle
\label{eq:bbp}
\ee
   
If we choose the Kerr cells so that $T= T'$, the phase into the probe due
to the photon in path 3 or 2 would be the same and the interference pattern 
${1 - \cos (\Theta) \over 2}$ is recovered for the signal field.

On the other hand, if one considers the case where the distance between
the two Kerr cells is larger than the coherence length of the probe laser
(e.g. 0.1 m for a single line argon laser) the two paths will still be
distinguishable (in general a random phase will appear between arm 2 and 3)
and interference will be lost.
For example, this scheme could  be realised taking a much shorter path
between BS I and Kerr Cell I than between BS II and Kerr II and then
compensating this difference after the Kerr cells, so as to have arm 2 and
3 equivalent.

This effect can be regulated, increasing the coherence length of the laser.
 In this way the interference pattern would thus be modulated by changing
the coherence length of the laser before injecting it into the first Kerr
cell. The observation of this effect represents, according to our opinion,
a very good and illustrative  example of the effect of disappearance of
quantum interference when {\it welcher Weg} information is obtained and of
the effect of erasing this information.
Furthermore, one can also think of erasing the quantum information after
the two Kerr cell (if decoherence effect are sufficiently tamed)  realising
a "true" quantum eraser, where also the delayed choice request is
implemented. Of course, single photons are necessary for the implementation
of the experiment, on the other hand the interference pattern will be
reconstructed on a whole ensemble of photons. Incidentally, it must be
acknowledged that, recently, an optical realisation of a delayed choice
quantum eraser, based on type II parametric down conversion, has been
realised \cite{Kim}.

Let us now investigate the possibility of a practical realisation of this
scheme.

Although admittedly very difficult,  the QND detection of a single photon
is at present possible \cite{SI,SM}. QND measurements of { \it welcher Weg}
have already been achieved 
using 100 meter long optical fibers (see Imoto et al. and Levenson et al.
\cite{QNDexp}). Of course, the implementation
of the present scheme using such devices would be, even though not impossible
in theory, almost impossible in practice. The
recent discovery of new materials with very high Kerr coupling, could
however permit an easier and more realistic,  implementation of this
experiment.

Two candidates as Kerr cell with ultra-high susceptibility to be used for
this scheme are the Quantum Coherent Atomic Systems (QCAS)  \cite{QCAS,SI}
and the Bose-Einstein condensate of ultracold (at nanoKelvin temperatures)
atomic gas \cite{BEC}.
These are recent great technical improvements which could permit the
realisation of small Kerr cells, capable of large phase shift, even with a
low-intensity probe. In fact, both exhibit extremely high
Kerr couplings compared to more traditional materials. In particular, the
QCAS is a rather simple system to be realised (for a review see
\cite{Arimondo})  and thus represents an ideal candidate to this role.
Incidentally, one can notice that Kerr coupling can be further enhanced by
enclosing the medium in a cavity \cite{Agarwal}.

In the following we will consider how we could set up this experiment using
the QCAS as Kerr cell.

Defining the Kerr coupling $\chi ^3$ through
\be
P_i(\omega) = 3 \epsilon_0 \Sigma_{j,k,l} \chi^3_{i,j,k,l} E_j(\omega)
E_k(\omega) E^*_l(\omega)  
\ee
where $P_i$ is the third order polarisation and $E_i$ are the electric
field components, the typical values of  $\chi ^3$  for traditional
materials (for example the optical fibers  used by Imoto et al. or Levenson
et al. \cite{QNDexp}), are around $10^{-22}-10^{-20} m^2 / V^2$ \cite{Chi3}.

For QCAS media much larger $\chi ^3$ can be obtained up to $10^{-8} m^2 / V^2$
\cite{SI,QCAS}, permitting large phase shifts with a short region of
interaction between the probe and the signal field. This allows for the
construction of a scheme where the probe field crosses both Kerr cells. The
optical distance between them can be kept of the order of 10cm - 1m. A
larger coherence length can be easily obtained, modulating it, for example,
by inserting different Fabry-Perot interferometers. 
It has been reported in the literature [9] that a 1 cm long Na cell can
provide a $10^{o}$ phase-shift with a photon flux as low as $10^9 $
photons/s. This number can be relaxed considerably (by at least two orders
of magnitude)  to the region of photon counting techniques, which, in the
visible part of the spectrum, can be considered the region below $10^7$
photons/second, which is roughly the limit of today detectors and related
electronics. 
Moreover, by substituting Na with Rb or Cs, different spectral regions can
be covered and this can be useful when using, for example, a parametric
fluorescence excited by an ultraviolet laser as the photon source (see below).

With respect to the above there seems to be a definite feasibility for
implementing such an experiment, although it should be recognised that the
practical problems are not  simple to overcome  and the existing technology
would be operated towards its limits.  
 
Next, we consider a possible extension of this experiment. Let us imagine
that we receive the first photon as a member of a polarisation entangled
state (the first photon is the input in our interferometer, while the
second will be photodetected at some other point D3, see fig.2); $H$ and
$V$ indicate horizontal and vertical polarisations 
\be
\vert \psi \rangle = { \vert H \rangle \vert V \rangle + \vert V \rangle
\vert H \rangle \over \sqrt {2}}
\label{eq:Psi}
\ee
as can be generated from a Type II down-conversion \cite{typeii} or from a
superposition of the parametric fluorescence produced in two Type I
crystals \cite{Hardy} or from one single Type I crystal using a beam
splitter \cite{typei}. 

Let us also use a polarising beam splitter to split the optical path on arm
2 and 3 (see fig.2): a photon with $H$ polarisation will follow path 2 and
a photon with $V$ polarisation will follow path 3. When the pump laser
crosses only the first Kerr cell, on arm 3, we obtain 

\be
\vert \Psi \rangle = { \vert H \rangle \vert V \rangle \vert \nu '\rangle +
\vert V \rangle 
\vert H \rangle \vert \nu \rangle \over \sqrt {2}}
\label{eq:Psi2}
\ee
where $\vert \nu \rangle$ denotes the probe field left unaffected by the
Kerr effect, while $ \vert \nu ' \rangle$ corresponds to the case where it
has interacted with the signal photon in the Kerr medium.
Incidentally, this state represents an interesting realisation of a
Schr\"odinger cat: a "macroscopic" state (the coherent field) is entangled
with the states of two microscopic systems (the two originally entangled
photons). How quickly decoherence will destroy this entanglement is beyond
the purposes of the present investigation.

If one considers the joint probability  $P(\theta _1, \theta _2)$ where the
photon is detected in D3, with a polariser at an angle $\theta _1$ with
respect to horizontal axis, in coincidence with the one in D1 or D2
(denoted with the index 2 in the following) , with a polariser at an angle
$\theta _2$ respect to the horizontal axis,  one obtains 
\bea
&P(\theta _1, \theta _2) = \langle \Psi \vert a^ {\dagger} _3 a^{\dagger}_2
a _2 a _3 \vert \Psi \rangle = 
 1/2  [ \cos^2 (\theta_1) \sin ^2 (\theta _2 ) 
+\cos ^2 (\theta_2) \sin ^2 (\theta _1 ) + \nonumber \\
& 2 \exp ( - 2 \vert \nu \vert ^2 \sin ^2 (\chi T) ) \cos [\vert \nu \vert
^2  \sin ( 2 \chi T) + \varphi] \cos (\theta_1) \sin  (\theta _2 ) \cos
(\theta_2) \sin  (\theta _1 )  ]
\label{eq:corr}
\eea
where $\varphi$ takes into account other contributions to the phase
differences aside from the one due to interaction with signal field in the
Kerr cell and 
where ideal detectors have been considered. Moreover, let in the following
$P(\Theta)$ represent the single detection probability.

When no probe field is inserted, these correlation functions lead to a
violation, for a proper choice of the polarizers, of the Bell  (or
Clauser-Horn) inequality, valid for local hidden variable theories. 
More in detail, for a maximal violation (selecting the proper angles for
polarizers) one has a value $CHS=0.207$ for the Clauser-Horn sum

\be
CHS = P(\theta _1, \theta _2) - P(\theta _1, \theta _2 ') + P(\theta _1 ',
\theta _2) + P(\theta _1 ', \theta _2 ') - P(\theta _1 ') - P( \theta _2) \,
\ee
while one would expect a negative value for a local theory. When the Kerr
effect modifies the phase of the probe field, this quantity is reduced and
becomes $\le 0$ when the second term in eq. \ref{eq:corr} is completely
suppressed [for the sake of brevity, we do not discuss here efficiency
loophole and the present state of Bell inequalities, see
\cite{typeii,Hardy,BI,Santos} and references therein]. Thus, one can test
the violation of the Bell inequality without probe field,  then when the
probe is inserted in the Kerr cell on arm 3 of the interferometer,  the
violation of Bell inequality will be reduced proportionally to 
$\Phi=\exp ( - 2 \vert \nu \vert ^2 \sin ^2 (\chi T) ) \cos [\vert \nu
\vert ^2  \sin ( 2 \chi T) + \varphi] $, which is related to the
signal-to-noise ratio in a homodyne measurement on the probe field $R =4
 \vert \nu \vert \sin (\chi T)$. For example, the maximum of $CHS$ is
reduced to $CHS=0.14$, $ 0.050$ and $=0.0025$ for $\Phi=0.8$, $0.5$ and
$0.1$ respectively. 

The violation will be reduced (to the point where it disappears), when the
set up  works with the probe laser in both cells, but with a distance
between the two cells larger than the probe coherence length. For the two
paths, when the photon propagates through  arm 2 ($H$ polarisation) or 3
($V$ polarisation) respectively, remain distinguishable (see the former
discussion). 

Finally, when the coherence of the pump field is increased, as explained
previously, the measurement of photon correlation functions will violate
the Bell inequality again, because now the two paths will be
indistinguishable  and one  will have a cancellation as in eq. \ref{eq:bbp}.

This experiment will therefore permit to relate the violation of Bell
inequality to obtaining or not the { \it welcher Weg} information.  
Of course, the effect on Bell inequalities is a direct consequence of the
destruction of interference pattern, however, we think that this last
example represents an interesting possibility to modulate decoherence on
entangled state and to quantify this effect by a measurement of CHS.

In summary, we have suggested an experiment where quantum interference is
modulated by the {\it welcher Weg} information obtained by a QND
measurement achieved in a Kerr cell. In detail, { \it welcher Weg}
information is obtained through the phase shift of a probe laser in two
different Kerr cells on the two different arms of an interferometer. The
probe crosses both the cells which have a relative distance larger than the
probe coherence length. The erasure is obtained thus by increasing the
probe coherence; for intermediate situations a partial restoration of
interference will be obtained.

We have also considered some practical details for the realisation of this
experiment, showing that it can be undertaken with  available technology.    

Finally, we have described how, using as the input in the former
interferometer a photon belonging to an entangled state, we can relate Bell
inequality violation to the { \it welcher weg} information erasure.

\vskip 1cm
{\bf Acknowledgements}

\noindent We would like to acknowledge support of ASI under contract LONO
500172. 

\noindent We thank E. Predazzi for useful comments about the manuscript.

\vskip 1cm

{

\end{document}